\documentclass[fleqn,usenatbib]{mnras}

\usepackage{newtxtext,newtxmath}

\usepackage[T1]{fontenc}

\DeclareRobustCommand{\VAN}[3]{#2}
\let\VANthebibliography\thebibliography
\def\thebibliography{\DeclareRobustCommand{\VAN}[3]{##3}\VANthebibliography}

\usepackage{graphicx}	
\usepackage{amsmath}	
\usepackage{balance}

\title[Clumps at cosmic noon with MICADO]{The MICADO first light imager for the ELT: Simulated Observations of Star Forming Clumps at Cosmic Noon}

\author[M. Simioni et al.]{Matteo Simioni$^{1}$\thanks{E-mail: matteo.simioni@inaf.it},
Carmelo Arcidiacono$^{1}$,
Andrea Grazian$^{1}$,
Marco Gullieuszik$^{1}$,
Elisa Portaluri$^{2}$,
\newauthor
Alvio Renzini$^{1}$,
Benedetta Vulcani$^{1}$,
Anita Zanella$^{1,3}$,
Francesca Annibali$^{3}$,
Michele Cirasuolo$^{4}$,
\newauthor
Richard Davies$^{5}$,
Natascha F\"{o}rster Schreiber$^{5}$,
Johanna Hartke$^{6,7,8}$,
Hanindyo Kuncarayakti$^{7,6}$,
\newauthor
Matteo Messa$^{3}$,
Stefan Raffetseder$^{9}$,
Letizia Scaloni$^{10,3}$,
Elena Valenti$^{4,11}$,
Eros Vanzella$^{3}$,
\newauthor
Jo\"{e}l Vernet$^{4}$,
and Roland Wagner$^{12}$\\
$^{1}$INAF - Osservatorio Astronomico di Padova, Vicolo dell'Osservatorio 5, Padova I-35122, Italy\\
$^{2}$INAF - Osservatorio Astronomico d'Abruzzo, Via Mentore Maggini, Teramo I-64100, Italy\\
$^{3}$INAF-Osservatorio di Astrofisica e Scienza dello Spazio di Bologna, Via Piero Gobetti 93/3,Bologna I-40129, Italy \\
$^{4}$European Southern Observatory, Karl-Schwarzschild-Strasse 2, Garching bei Munchen D-85748, Germany\\
$^{5}$MPE - Max-Planck-Institut für extraterrestrische Physik Giessenbachstrasse 1, Garching D-85748, Germany\\
$^{6}$Finnish Centre for Astronomy with ESO, (FINCA), University of Turku, Turku FI-20014, Finland\\
$^{7}$Tuorla Observatory, Department of Physics and Astronomy, University of Turku, Turku FI-20014, Finland\\
$^{8}$Turku Collegium for Science, Medicine and Technology (TCSMT), University of Turku, FI-20014 Turku, Finland\\
$^{9}$Industrial Mathematics Institute, Johannes Kepler University (JKU),Altenberger Stra{\ss}e 69, Linz A-4040, Austria \\
$^{10}$Department of Physics and Astronomy “Augusto Righi”, University of Bologna, Via Piero Gobetti 93/2,Bologna I-40129, Italy\\
$^{11}$Excellence Cluster ORIGINS, Boltzmann-Strasse 2,Garching Bei Munchen D-85748, Germany\\
$^{12}$Johann Radon Institute for Computational and Applied Mathematics of the Austrian Academy of Sciences (RICAM), Altenberger Stra{\ss}e 69, Linz A-4040, Austria\\
}
\date{Accepted 2026 September 01. Received 2026 August 09; in original form 2026 June 01}

\pubyear{\the\year{}}

\begin{document}
\label{firstpage}
\pagerange{\pageref{firstpage}--\pageref{lastpage}}
\maketitle

\begin{abstract}
Galaxies at redshift z$\sim$2-6 exhibit prominent star-forming regions (clumps).  Characterisation of these galactic structures requires both high sensitivity and spatial resolution of $\sim$100$\,$pc or less. Currently, these spatial scales are achievable only with the aid of strong gravitational lensing. However, lensing introduces model-dependent uncertainties and limits the sample to galaxies behind massive clusters. The upcoming 40m-class telescopes, such as the ESO Extremely Large Telescope (ELT), will enable routine studies of clumps in ubiquitous, non-lensed z$\geq$2 galaxies.
We assess the capability of MICADO, the first-light imager and spectrograph of the ELT, to characterise clumps in non-lensed z=2 galaxies. Specifically, we focus on two representative observing scenarios, providing the basis for an early scientific exploitation of the instrument.
We modelled clumps in an idealised z=2 star-forming galaxy and produced mock MICADO observations in both broad- and narrow-band imaging. The latter leverages the strong emission lines typical of clumps to improve detection and characterisation.
Reaching UV rest-frame magnitudes of clumps as faint as $\rm M_{UV}\sim-15$ in optimal observing conditions, MICADO will be able to efficiently characterise clumps in non-lensed z=2 galaxies down to ${\rm R}_{\rm e}$$\sim$20$\,$pc. Moreover, the full MICADO $\sim$1$\,$arcmin$^{2}$ field of view enables efficient surveys of multiple targets.
This study demonstrates MICADO’s potential to fill a crucial observational gap, shedding new light on clump formation and evolution at cosmic noon. Accurate PSF reconstruction will be crucial to fully harness this capability as it significantly impacts clump detection and characterisation, especially at the smallest sizes (<40$\,$pc).
\end{abstract}

\begin{keywords}
galaxies: photometry, galaxies: high-redshift, galaxies: star clusters: general,
infrared: galaxies, instrumentation: adaptive optics, techniques: high angular resolution
\end{keywords}



%
\section{Introduction}\label{sec:intro}
It has been known for over two decades that galaxies at $z > 1$ have more irregular UV morphologies than their local counterparts and host active sites of star formation, typically dubbed ``clumps" \citep{Elmegreen2008, Genzel2011,ForsterSchreiber2011, Conselice2014, Guo2015, Zanella2015, Huertas-Company2023}.
Clumps in z$\sim$1-3 galaxies are thought to form as a result of the fragmentation of gas-rich galactic disks driven by the Toomre Instability \citep{Toomre1964}, as discussed e.g, by \cite{Binney2008, Genzel2011, Livermore2012, Ceverino2012, Reina-Campos2017, Zanella2021}. Observations appear to be largely supportive of this scenario \citep{Shibuya2016, Guo2018, Zanella2019, Zanella2024}. In gravitationally unstable disks, the resulting clumps are expected to have masses and sizes that depend on the surface density of the host galaxy \citep{Livermore2012, Ceverino2012, Reina-Campos2017, Zanella2021}. However, observational constraints on clump masses, sizes, and ages remain relatively poor. The Hubble Space Telescope (HST) has characterised the rest-frame UV and optical properties of clumps revealing typical stellar masses of $\mathrm{M_\star \sim 10^7 - 10^9\, M_\odot}$, star formation rates of $\mathrm{SFR \sim 0.1 - 10\, M_\odot\, yr^{-1}}$, and mostly spatially unresolved sizes $< 1\, \mathrm{kpc}$ \citep{Wuyts2013, Guo2018, Zanella2019, Sattari2023}. More recently, observations with the  James Webb Space Telescope (JWST) have enabled the study of clumps in the infrared (up to $\sim 5\;\mu$m), a previously uncharted wavelength regime, with a spatial resolution of $\sim 0\farcs15$. At shorter wavelengths ($\sim 1-2\;\mu$m), JWST also provides a spatial resolution nearly three times better than HST.
These observations have confirmed that clumps typically contribute $1 - 30\%$ of the total star formation rate of their host galaxies, while accounting for $1 - 20\%$ of their stellar mass \citep{Kalita2023}. 

Many of the current studies focusing on high-redshift clumps take advantage of gravitational lensing, which amplifies and magnifies the light coming from distant objects, making them much easier to observe and analyse. Thanks to this technique, clumps with sizes of tens to hundreds of pc are now routinely detected \citep{Jones2010, Adamo2013, Cava2018, Zick2020, Vanzella2022,  Mestric2022, Messa2022, Bolamperti2023, Claeyssens2023, Vanzella2023, Adamo2024, Fujimoto2024, Mowla2024, Messa2024, Messa2025, Messa2026, Claeyssens2025}. These studies have also resolved large clumps into clusters of smaller ones, as expected from simulations suggesting that massive clumps of size $\sim 1\,\mathrm{kpc}$ would be resolved into smaller star-forming complexes if observed at sufficiently high spatial resolution \citep{Mayer2016, Behrendt2016}.

Comparing lensed and non-lensed clumps is challenging, as the high resolution provided by lensing surpasses the resolution limits of present non-lensed studies \citep[limited to sizes down to $\rm{R_e} \sim 0.5 -1\, {\rm kpc}$; e.g.][]{Ji2024,Kalita2024,MarquesChavez2024,Kalita2025a,Kalita2025b}. This limitation may result in systematic overestimation of clump sizes and stellar masses in non-lensed samples due to unresolved blending \citep{Tamburello2015, Fisher2017, Huertas-Company2020}.

Obtaining accurate sizes and masses for the clumps also has important implications for understanding their survival and ultimate fate. In fact, there is an ongoing debate on whether clumps can survive for several hundred Myr, migrate inward, and contribute to the formation of the bulge and the structural evolution of the host galaxy \citep{Elmegreen2007, Mandelker2014, Mandelker2017}. In contrast, other studies argue that clumps are relatively low in mass and small in size, surviving only 50 to 100 Myr, and therefore may play a smaller role in bulge formation \citep{Murray2010, Genel2012, Hopkins2012, Hopkins2014, Buck2017, Oklopcic2017}.

While studies of lensed systems are exceptionally powerful for probing the distant Universe, they require highly precise magnification modelling to robustly disentangle the intrinsic properties of the background sources from lens-induced distortion. Furthermore, the sample size of strongly magnified objects (${\rm \mu}\geq10$) remains severely limited by the intrinsic rarity of such alignments. This observational bias is especially pronounced for the most massive clumps ($>~10^9\; {\rm M}_\odot$), which are under-represented in current lensing studies (see Fig. 17 in \citealt{Kalita2025a}). As discussed in the aforementioned work, the lack of massive ($>~10^9\; {\rm M}_\odot$) clumps in lensed systems suggests that these objects are the result of the spatial blending of multiple, less massive sub-clumps. While gravitational lensing currently provides the spatial resolution needed to resolve these sub-units individually, they would appear as a single, more massive clump in non-lensed galaxies. Therefore, it is crucial to complement lensing studies by observing high-redshift clumps with the unprecedented spatial resolution enabled by adaptive optics (AO) on upcoming 40-m class ground-based telescopes. 

In this respect, the European Southern Observatory's Extremely Large Telescope \citep[ELT,][]{elt2007,elt2008} will be an absolute game changer, as it will deliver the sharpest view of the Universe ever achieved in infrared bands. Since its first light, currently planned for the end of 2030, it will be equipped with the Multi-AO Imaging Camera for Deep Observations (MICADO; \citealt{2016SPIE.9908E..1ZD}) that will provide almost diffraction-limited near-infrared observations thanks to state-of-the-art adaptive optics (AO) technology. This results in an expected resolving power that can be up to $6.5$ times better than that of JWST \citep{Gardner2006}. 
 
Future  MICADO observations hold the potential to fundamentally address key open questions in clump physics, such as their size distribution, internal substructures, mass and star formation densities, and their role in funnelling radial gas flows that promote galactic bulge growth. Crucially, accurately measuring the physical sizes of these systems is a prerequisite  for these studies, as it disentangles 
whether clumps are monolithic structures or complex assemblies of smaller components.  
To quantitatively evaluate MICADO's performance, in this paper, we simulate non-lensed star-forming clumps at $z\sim 2$ and assess the instrument's capability to retrieve their basic photometric and structural properties, specifically magnitude and angular size. We investigate the angular scales and depths achievable with K-band observations and compare our predictions with current state-of-the-art JWST observations.

The paper is structured as follows. In Section~\ref{sec:micado} we briefly outline the baseline performance of MICADO relevant to high-redshift clumps science. A detailed description of our simulation strategy and the sampled parameter space is provided in Section~\ref{sec:simu}. The results are presented in Section~\ref{sec:res}, and further discussed in Section~\ref{sec:discussion}. Finally, our conclusions on the future exploitation of MICADO for this science case are summarised in Section~\ref{sec:wup}.

Throughout the paper, we adopt a concordance cosmology with ${\rm H}_0 = 70\, {\rm km~ s^{-1} Mpc^{-1}}$, $\Omega_{\rm M} = 0.3$, and $\Omega_\Lambda$ = 0.7. All magnitudes are in the ABmag system, if not otherwise stated.


\section{MICADO in a Nutshell}\label{sec:micado}
During its first operational phase, MICADO will rely on single-conjugated AO (SCAO), implying that its performance will vary over its $50.5$\arcsec$\times50.5$\arcsec field of view (FoV). Specifically, the Strehl ratio (SR) is expected to degrade as the distance from the AO reference star increases, from a limiting $\sim60\%$ for a target close to the AO natural guide star, down to $\sim12\%$ some 30\arcsec from it.
The performance will improve when the Multiconjugate Adaptive Optics Relay For ELT Observations (MORFEO; \citealt{ciliegi_maory_2020,2022SPIE12185E..14C}) will be integrated (in 2032 according to the current schedule): a multi-conjugate AO \citep[MCAO,][]{becker88,beckers89a} correction will ensure almost constant performances over the entire FoV, with relaxed requirements concerning the natural AO reference stars. In practice, MORFEO+MICADO is expected to routinely reach a SR$\simeq 40\%$, approaching $\sim 60\%$ under optimal conditions in K band. 
Unless otherwise stated, we will focus on the K band, where AO correction is more effective because the SR increases with wavelength for a given residual wavefront error.

\begin{figure}
    \centering
    \includegraphics[width=0.99\linewidth]{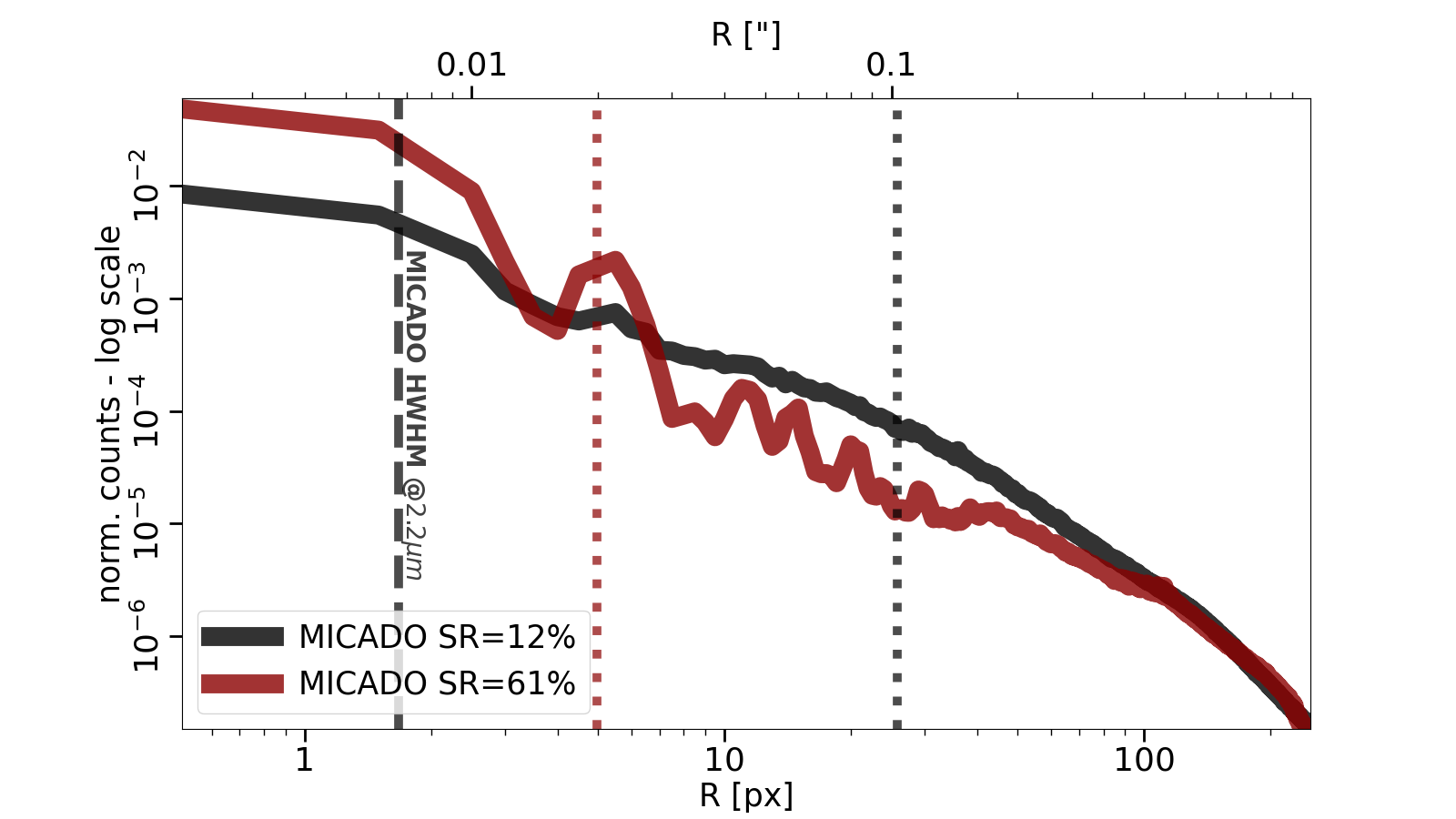}
    \caption{Comparison between the radial profile of the MICADO SCAO 30" off-axis PSF (black solid line) and the MCAO MICADO+MORFEO PSF (red solid line). Notice that, while the two profiles have been normalised to have the same integral, the SR in the SCAO mode ($30"$ off-axis) is $12\%$, much lower than SR$\sim60\%$ obtained in the MCAO mode. All scales are logarithmic. Both PSFs have similar FWHM values, indicated by the vertical dashed line (which marks the HWHM), but significantly different half-light radii (dotted vertical lines; colours are consistent with the legend). With SR$=61\%$, light is more concentrated in the core of the PSF, resulting in enhanced sensitivity.}
    \label{fig:scaovmcao}
\end{figure}
An example of the two most representative MICADO point spread functions (PSFs), which will be used throughout this work, is shown in Figure~\ref{fig:scaovmcao}. As a preview of what detailed in Section~\ref{sec:simu_obs}, the PSF with SR$=12\%$ (black solid line) represents the SCAO PSF performance. This profile corresponds to typical observing conditions at an off-axis distance of R=$30$\arcsec from the AO reference star, serving as a conservative baseline for first-light MICADO observations. In contrast, the PSF with SR$=61\%$ (red solid line) illustrates the capabilities of MICADO+MORFEO MCAO under optimal conditions (specifically, an ideal guide star asterism and good atmospheric conditions).
The $39$m diameter of the primary ELT mirror, combined with AO correction, will produce an extremely narrow PSF, around $13{\rm mas}$ of full-width at half maximum (FWHM) in K band, making this instrument particularly suitable for high spatial-resolution observations of faint and compact targets, such as star-forming clumps. 
In particular, the two PSFs in Fig.~\ref{fig:scaovmcao} have very similar FWHM values: the black dashed, vertical line marks the half-width at half maximum (HWHM). But light is significantly more concentrated in the case of SR=$61\%$, as shown by the half-light radii (dotted vertical lines).

The MICADO consortium will provide users with a PSF reconstruction (PSF-R) service that will generate template PSFs, specific to each selected position within the FoV, for each exposure \citep{2020SPIE11448E..37S,2022SPIE12185E..41G}. A key aspect of the adopted PSF-R strategy is that the reconstruction is performed without relying on information coming from the scientific frames (i.e. the actual observations). Instead, the reconstruction will rely on AO telemetry \citep{2018JATIS...4d9003W}. This is particularly indicated in the case of extragalactic targets like the ones considered in this work. Indeed, PSF-R will avoid the necessity of including within the FoV suitable point sources (in position, brightness, and number) to build a reliable PSF template, as this would greatly limit the observable sky, as well as introducing artefacts and additional background noise that may affect the detection and characterisation of the targets.

\section{Simulations}\label{sec:simu}
Our goal is to assess the observability of compact star-forming regions in $z\sim2$ galaxies. To this aim, since we are mainly interested in studying the intrinsic observability of these clumps, we excluded the subtle systematics arising from real observations, such as crowding effects and biases induced by the specific measurement technique used (see Sec.~\ref{sec:scieval}).

Expanding the work of \citet{2016A&A...593A..24G}, and following the method described in \citet{2022JATIS...8c8003S}, we simulated a set of high-redshift clumps using GALFIT \citep{2010AJ....139.2097P}. We vary both the intrinsic size and brightness of the clumps while fixing the local background, representing the host galaxy. This approach allows some flexibility while keeping the number of free parameters to an acceptable level.

Next, we simulate MICADO wide-field mode (pixel scale 4mas/pix) observations in the SCAO mode with SimCADO \citep{2016SPIE.9911E..24L,2020SPIE11452E..1ZL}, adding the effect of instrument PSF and the various noise contributions. The MCAO case has also been considered, assuming optimal observing conditions.

Operating in the near-infrared ($0.8-2.45\mu{\rm m}$), it is possible to sample the rest-frame optical continuum of our targets ($\sim3000-8000$\AA). In addition, the ${\rm Br}-\gamma$ narrow-band filter (width $289$\AA) targets the ${\rm H}{\rm \alpha}$ emission (rest-frame $\lambda=6562.8$\AA) in a narrow redshift range around z$=2.31$. We simulate observations in both the K band and the ${\rm Br}-\gamma$ narrow-band filter: the K band simulations assess the detectability, size measurement, and substructure of clumps, while the narrow-band simulations provide estimates of clump star formation rates via ${\rm H}{\rm \alpha}$ emission \citep{2012ARA&A..50..531K}. The latter can be estimated by the excess flux in the narrow-band filter over the continuum sampled by the K band.
In the following, we will go through each step of the simulation process.

\subsection{Star-forming clumps}\label{sec:host_gal}
We model the clumps as circular 2D Gaussian profiles, running several simulations varying the clump properties. A total of $144$ unique combinations of clump sizes and luminosities have been used, sampling the range observed for real clumps by e.g., \citet{Mestric2022} and \citet{Claeyssens2023}. In all cases, a single clump is simulated, overimposed on a surface brightness of constant value, of $\mu_{\rm K, Reff,host}=20\,{\rm mag/arcsec^{2}}$. This is equivalent of placing the clump at a radial distance of the effective radius of a typical star-forming z$\sim2$ galaxy. A detailed justification for this choice is provided at the end of this section.

\begin{table}
    \centering
    \caption{\label{tab:simu_range} Range of parameter values for the simulated clumps. The redshift of the target has been fixed at z$=2.31$. Both the case of SR$=12\%$ and SR$=61\%$ are considered. We list the absolute magnitude in the rest-frame UV $({\rm M}_{\rm UV})$, the effective radius $({\rm R}_{\rm e})$, sampled uniformly in logarithmic scale, and the apparent integrated magnitude range used for simulating narrow-band observations. The number of samplings is also indicated.}
    \begin{tabular}{llll}
    Parameter                   & SR     & Range               &samplings \\
    \hline
    ${\rm M}_{\rm UV}$          & $12\%$ & $[-19.3,-16.8]$ mag & $6$      \\
                                & $61\%$ & $[-19.3,-14.8]$ mag & $10$     \\
    ${\rm R}_{\rm e}$           & $12\%$ & $[4,\; 778]$ pc     & $24$     \\
                                & $61\%$ & $[4,\; 778]$ pc     & $24$     \\
    ${\rm m}_{{\rm Br}-\gamma}$ & $12\%$ & $[21.8,27.3]$ mag   & $12$     \\
                                & $61\%$ & $[24.3,29.8]$ mag   & $12$     \\
    \end{tabular}
\end{table}
In Table~\ref{tab:simu_range} we list the selected value range for all the relevant clump parameters, in the case of SCAO and MCAO observations.
The size of the clumps (in this case the effective radius, ${\rm R}_{\rm e}$) is given $24$ values evenly spaced on a logarithmic scale in the interval $[4,\; 778]$ pc, which corresponds to ${\rm R_{e}}\in [0.5,\; 93]\, {\rm mas}$, or equivalently to ${\rm R_{e}}\in [0.12,\; 23]\, {\rm px}$, at z$=2.31$.
In the SCAO case, for each size, we then explore an interval of $2.5$ mag, sampling $6$ magnitude values linearly distributed in the UV rest-frame $-19.3\,{\rm mag} \leq {\rm M}_{\rm UV} \leq -16.8\,{\rm mag}$ range. In the case of MCAO observations, we adopt the same clump size range as for SCAO, but extended the range of magnitude down to $2\,{\rm mag}$ fainter ($-19.3\,{\rm mag} \leq {\rm M}_{\rm UV} \leq -14.8\,{\rm mag}$). This is to take into account the enhanced performance of the instrument, given the higher resulting SR in this case. The bin size in magnitude has been kept constant; consequently, the total number of samplings in magnitude has risen to $10$ in the case of MCAO simulated observations.
We simulate each combination of clump size and luminosity $120$ times to have statistically significant results.

To convert the integrated absolute UV rest-frame magnitudes of the clumps into apparent ones at ${\rm \lambda}=2.2{\rm \mu m}$, we assume a flat spectral energy distribution for the clumps \citep{Mestric2022,Bolamperti2023,Claeyssens2023,Claeyssens2025}. 

To define the magnitude range for the narrow-band case, we first convert the clump star formation rate (SFR) from 
\citet{Mestric2022} and \citet{Claeyssens2023} into ${\rm H\alpha}$  luminosity (${\rm L}_{\rm H\alpha}$) using the relation of \citet{1998ARA&A..36..189K}
and then convert ${\rm L}_{\rm H\alpha}$ in flux density and ultimately (${\rm f}_{{\rm H\alpha,\lambda}}$), accounting for the  ${\rm Br}-\gamma$ filter.
Finally, we convert ${\rm f}_{{\rm H\alpha,\lambda}}$ into AB magnitudes.
No extinction has been considered, since for young star-forming regions at cosmic noon, it is expected to be relatively low on average. For completeness, in the case of \citet{Claeyssens2023}, $50\%$ of their clumps have ${\rm E(B-V)}\leq0.1$ ($40\%$ have ${\rm E(B-V)}=0$), which corresponds to about ${\rm \Delta}_{\rm mBr-\gamma}=0.3\,$mag if the \citet{Calzetti2000} extinction law is applied.

The resulting magnitude intervals are listed in the last row of Table~\ref{tab:simu_range}. We selected $12$ linearly distributed and evenly spaced values. For MCAO, we shifted the interval by $2.5 {\rm mag}$.
In total, considering both broad- and narrow-band observations, we generated and analysed 115200 simulated clumps.

\begin{figure*}
    \centering
    \includegraphics[width=\textwidth]{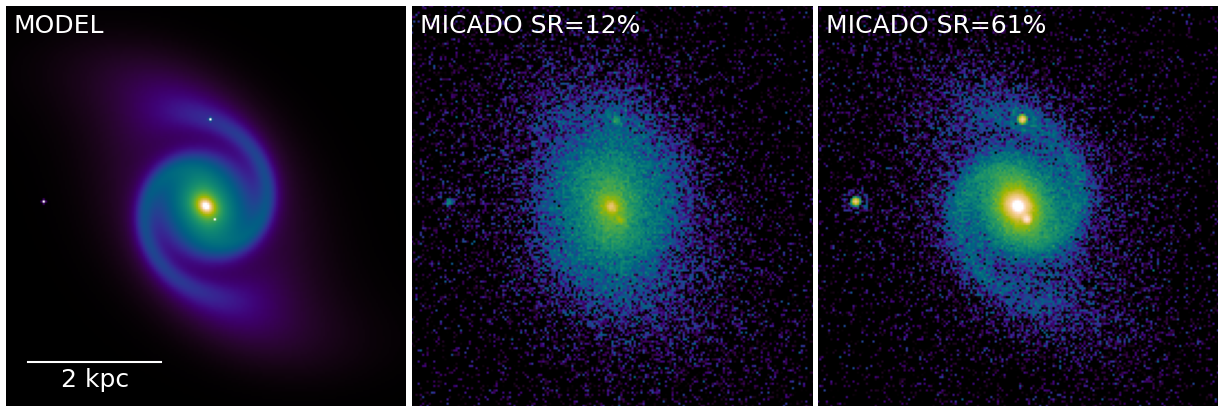}
    \caption{Effect of different sub-structures of the host galaxy on clump detectability. {\em Left panel:} GALFIT model of a z=2 galaxy. For illustrative purposes, three clumps are included in the model, placed at increasing radial distances. The inner one is placed near the core of the host galaxy, $200\, {\rm pc}$ from it. {\em Middle panel:} The resulting simulated SCAO observations. The only detectable feature of the galaxy is, as expected, the core. The clumps luminosity has been set to be at the detection limit for this MICADO configuration, identical for all of them. {\em Right panel:} MICADO+MORFEO MCAO simulated observation. The effect of the enhanced sensitivity can be appreciated. It is due to the increase in PSF SR, with respect to the SCAO configuration. In both cases, we considered a total exposure time of 3 hours in the K band. The galaxy features, even the brightest ones, do not significantly affect the observability of the clumps in either case. The main reason is the high expected background, combined with the extremely small MICADO pixel scale ($4\,{\rm mas/px}$ in wide mode). Even in the case of SR$=61\%$, where the most compact, bright features of the host galaxy are detectable, their contribution to the local noise level is negligible in comparison to the background.}
    \label{fig:simobs}
\end{figure*}

As anticipated, we accounted for the effect of the host galaxy, just by adding its contribution as a constant surface brightness to the simulated images, derived from detailed simulations of typical z$=2$ clumpy galaxies (Scaloni et al. in prep.). Figure~\ref{fig:simobs} shows the host galaxy model that we produced to estimate the background to be added to the clump simulations (left panel). The galaxy is modelled as a typical star-forming galaxy at ${\rm z}\sim2$ with aself control bulge, spiral arms and a disk, yielding an effective radius of $1.25{\rm kpc}$ (corresponding to $150{\rm mas}$ at z$=2.31$, or $37.5{\rm px}$ in the MICADO wide mode). Using the mass-luminosity relation from \citet{2014ApJ...788...28V} as a reference, we assigned the galaxy an integrated magnitude of ${\rm m}_{gal,K}=22\,{\rm mag}$ (about ${\rm log[M_\ast/M_{\odot}]}=10.25$). To evaluate the host galaxy's contribution to the local background, we estimated its surface brightness at three reference locations. 
The first is in the central core region, which exhibits a surface brightness of $\mu_{\rm gal,core,K}=17\,{\rm mag}/{\rm arcsec}^2$. The second region was selected at the effective radius, along the spiral arm, resulting in a local surface brightness of $\mu_{\rm gal,Re,K}=20\,{\rm mag}/{\rm arcsec}^2$. The third region is located at $2.5\,{\rm kpc}$ ($2$ effective radii), where the local surface brightness drops to $\mu_{gal,2Re,K}=22\,{\rm mag}/{\rm arcsec}^2$. The central panel of Figure~\ref{fig:simobs} displays the corresponding MICADO SCAO simulated observation ($3\,{\rm hr}$ total exposure time in K band, see below), where the bulge is the only clearly visible feature of the host galaxy. The simulated MCAO observation is shown in the right panel; in this configuration, most of the compact structures of the host galaxy become detectable. 
For completeness, we also included three clumps in Figure~\ref{fig:simobs}, modelled as point sources with a magnitude equal to the detection limit for K-band SCAO observations (${\rm m}_{\rm K}\simeq28\,$mag). To avoid confusion with the galaxy core, the innermost clump was placed at a physical distance of $200\, {\rm pc}$ from the centre. A visual inspection allows for a qualitative assessment of how the varying local backgrounds across these three selected regions affect clump observability. In both SCAO and MCAO cases, the clumps remain detectable. Specifically, in the dense central region, the associated signal is just $10\%$ of the local host galaxy background. This additional background flux increases the local noise by $5\%$, leading to a reduction in the limiting magnitude of approximately $0.1\,{\rm mag}$. For comparison, the signal from the host galaxy accounts for only $2.5\%$ and $0.3\%$ of the background in the other two regions, respectively

We conclude that the effect of the host galaxy on clump observability is negligible due to two primary factors. First and foremost is the high background expected for ground-based observations in the K band. While telluric line emission is relatively low in the K band compared to the H or J bands, ground-based near-infrared observations are heavily dominated by the thermal emission of the telescope itself, which typically reaches 13 ${\rm VEGAmag/arcsec^2}$ \citep{2000Msngr.101....2C}. 
The second factor is MICADO's extremely fine pixel-scale ($4\,{\rm mas/px}$ in wide mode).  While this small scale is highly advantageous for reducing the background flux collected per pixel, thereby lowering the associated photon noise for compact clumps, it similarly dilutes the emission of any extended source by spreading its signal over a large number of pixels, ultimately mitigating its local impact on clump detection.

\subsection{Simulated observing strategy}\label{sec:simu_obs}
\begin{table}
    \centering
    \caption{\label{tab:psf_prop} Properties of the used MICADO PSFs. The listed parameters include SR, full-width at half maximum (FWHM), encircled energy of the core (${\rm EE}_{\rm C}$), and half-light radius (${\rm R}_{50}$), all these values for the K band filter.}
    \begin{tabular}{l|r|r}
                       & MICADO SR$=12\%$ & MICADO  SR$=61\%$ \\
    \hline
    SR                 & $12\%$           & $61\%$            \\
    FWHM               & $13.4$ mas       & $13.3$ mas        \\
    ${\rm EE}_{\rm C}$ & $7\%$            & $31\%$            \\
    ${\rm R}_{50}$     & $103$ mas        & $20$ mas          
    \end{tabular}
\end{table}
As mentioned in Sec. 2, in the case of SCAO observations, the quality of the PSF degrades with increasing distance from the AO reference star ($\theta$) and with decreasing isoplanatic angle ($\theta_0$). 
The SR measures the quality of  PSF and scales as \citep{1982JOSA...72...52F,1993ARA&A..31...13B,1999aoa..book.....R,2005A&A...438..757C}:
\begin{equation*}
    {\rm SR}(\theta) \propto exp \left(-\left(\frac{\theta}{\theta_0}\right)^{\frac{5}{3}}\right)\,.
\end{equation*}
The exponential relationship shows how quickly the quality of the PSF deteriorates moving away from the reference star (the isoplanatic patch). 
The isoplanatic angle, which defines the size of this region,  is related to the Fried parameter (that describes the turbulence in the atmosphere) and increases with longer wavelength ($\theta_0\propto\lambda^\frac{6}{5}$). By observing at longer wavelengths, good PSF quality can be maintained over a larger area around the AO reference star, effectively expanding the corrected field. For instance, by using the K-band, which we adopt here, the area where ${\rm SR}\geq10\%$ extends to about 0.8 arcmin$^2$ around each AO reference star (corresponding to radial distances of $30\arcsec$). This area is comparable to the MICADO FoV.

The SCAO PSF, used in this work, has been produced by the MICADO consortium\footnote{Produced in the context of the Final Design Review of the instrument.}. 
This PSF was generated under typical Paranal atmospheric conditions and obtained using a point-like AO reference source in the intermediate-brightness regime ($\rm m_{R}=14$ VEGAmag). The adopted PSF is characterised by an ${\rm SR}>60\%$ for the on-axis direction. At $30\arcsec$ off-axis distance, the SR drops to $12\%$; however, the PSF retains an almost diffraction-limited core \citep{Clenet2014}.  
If the AO reference star is positioned at the centre of the MICADO detector, an area of $0.18\, {\rm arcmin}^2$ ($24\%$ of the FoV) will have SR$>40\%$. Expanding the threshold to SR$>20\%$ increases the usable area to $0.52\,{\rm arcmin}^2$ ($67\%$ of the FoV), and further relaxing the threshold to SR$>10\%$ increases it to $0.73\, {\rm arcmin}^2$ ($94\%$ of the FoV).

On the other hand, the MCAO PSF has been obtained using PASSATA \citep{Agapito2016}. As mentioned in Section~\ref{sec:micado}, extremely good atmospheric conditions have been assumed, specifically a Q1 turbulence category, with a seeing of 0\farcs43 at 500nm (${\rm r}_0=0.234\,$m and ${\rm t}_0=8.08\,$ms). This, coupled with a favourable asterism composed of a mix of laser and natural guide stars, provides the high-performance MORFEO case considered in the present work.

For completeness, we report in Table~\ref{tab:psf_prop} a quantitative description of the two PSFs adopted in this work, and already presented in Section~\ref{sec:micado} (their radial profiles being shown in Figure~\ref{fig:scaovmcao}). A classic PSF metric is used, defined by SR, FWHM, encircled energy in the PSF core (${\rm EE}_{\rm C}$) and half-light radius (${\rm R}_50$).

We simulate observations of 3~hr, dividing the total exposure time into $180$ single exposures of $60$ s each. This detector integration time (DIT) strikes a good balance, allowing us to handle the high background typical of K filter observations. Integrating over $180$ DITs ensures a good limiting magnitude and a relatively stable PSF. For K filter observations, doubling the total integration time would yield a gain of half a magnitude. 
For narrow-band ${\rm Br}-\gamma$ observations, we adopt the same DIT and total integration time as for the K band, despite the lower predicted sky background, which decreases from $13$ to $15.7$ VEGAmag/arcsec$^2$. 

\section{Results}\label{sec:res}
\subsection{Broad-band Observations}
\begin{figure}
    \centering
    \includegraphics[trim={3cm 1.5cm 1.5cm 1.5cm},width=0.9\linewidth]{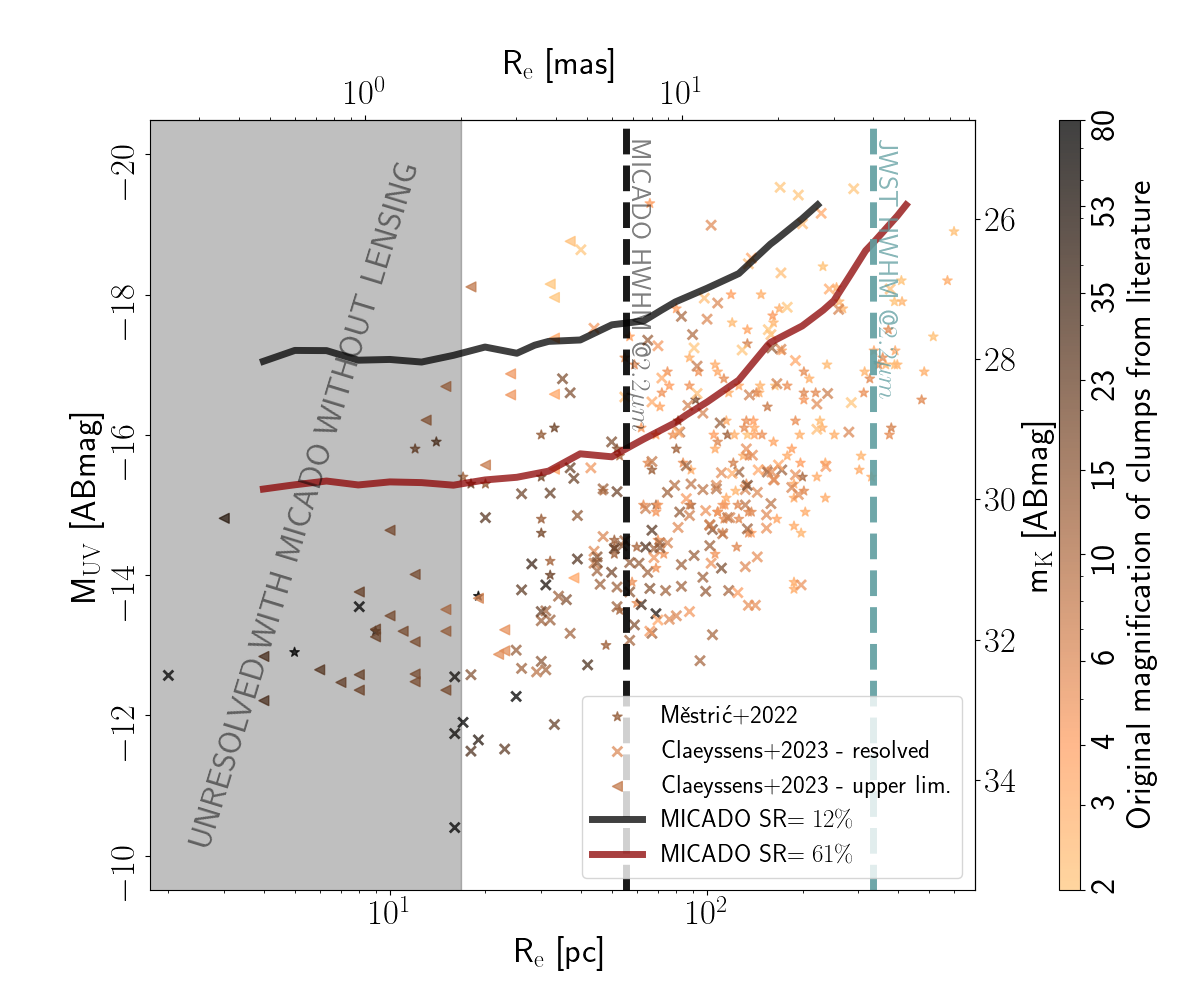}
    \caption{Observability of non-lensed, compact star-forming clumps at z$\sim2$ in the M$_{\rm UV}$-size plane for broad-band K observations.  Literature results for clumps in gravitationally lensed galaxies are shown as stars (from 
    \citealt{Mestric2022}) and crosses (from 
    \citealt{Claeyssens2023}, left-pointing triangles for upper limits), colour-coded by their measured gravitational magnification factor. The black solid curve shows the SNR$=5$ threshold for $3\,$hr MICADO SCAO observations in the K band, assuming a conservative $30\arcsec$ 
    off-axis PSF (SR$=12\%$). Clumps located above this line are considered detectable. For comparison, the red solid curve shows the threshold for MCAO observations (MICADO+MORFEO, SR$=61\%$, see Section\ref{sec:simu_obs}) for the same integration time. Dashed vertical lines indicate the PSF HWHM for MICADO (black) and JWST (azure), highlighting the gap in spatial resolution between the two facilities. While a well-characterised PSF enables reliable clump size measurements down to $R_e$ on the order of a fraction of a pixel, clumps falling within the grey-shaded area remain fundamentally unresolved, even with MICADO.}
    \label{fig:resk}
\end{figure}
The unprecedented capability of MICADO to resolve high-redshift star-forming clumps without resorting to gravitational lensing is demonstrated in Figure~\ref{fig:resk}, which presents the main result of our work: the inferred properties (UV magnitude and size) of star-forming clumps in non-lensed $z\sim2$ galaxies as observed by MICADO using the broad K-band filter. The black solid line indicates the signal-to-noise ratio (SNR) limit for MICADO SCAO observations at an off-axis distance of $30\arcsec$, with the SNR set to $5$. For comparison, the red line represents the SNR$=5$ level expected from MCAO observations. In both cases, the total exposure time and wavelength range are kept the same. 

For all configurations, the SNR has been computed adopting a circular area of $5\,$px in diameter to approximately sample the PSF core in MICADO wide mode. This choice maximises the SNR for compact clusters, while systematically underestimating the flux from the extended structures. Furthermore, for clump sizes significantly larger than the MICADO PSF, the native MICADO wide mode pixel scale ($4$ mas/px) may become sub-optimal, and spatial binning could be considered. A detailed exploration of these effects is beyond the scope of the present work.

The dashed vertical lines in Figure~\ref{fig:resk},  mark the PSF HWHM for MICADO (black) and  JWST (azure), added for comparison. The separation between these lines directly reflects the primary mirrors diameter relative ratio of the two facilities. While a well-characterised PSF enables reliable clump size measurements down to effective radii ($R_e$) on the order of a fraction of a pixel, clumps falling below this hardware threshold remain fundamentally unresolved even with MICADO, unless boosted by strong-lensing magnification.

To place these performance limits into a broader observational context, we compare our predictions with existing high-redshift clump measurements from the literature. Specifically,  Figure~\ref{fig:resk} includes intrinsic, de-lensed data points from two recent lensing studies
 \citep{Mestric2022, Claeyssens2023}, with symbols colour-coded by their measured gravitational magnification. From this comparison, it is evident that a magnification $\mu\geq5$ is typically required to resolve and measure the sizes of the most compact clumps with JWST. Conversely, the figure clearly shows that MICADO can access the brightest non-lensed clumps (${\rm m}_{\rm K}<28-30\,$mag) in blank fields, resolving their physical sizes down to approximately $20\,$pc, corresponding to just under one wide-mode pixel, without the aid of gravitational lensing. We emphasise that while relying on strong gravitational lensing inherently reduces the number of available targets and introduces additional uncertainties into the measurement error budget,  it provides a substantial gain in both sensitivity and spatial resolution. A detailed treatment of these trade-offs is beyond the scope of the present work; a dedicated study on the synergy between MICADO and strong gravitational lensing is currently in preparation (Messa et al. in prep). Nevertheless, for completeness, a qualitative assessment of MICADO's performance under a magnification factor of ${\rm \mu}=10$ yields a $2.5\,$mag deeper limiting magnitude (reaching ${\rm M}_{\rm UV}\sim-13\,$mag for the most compact clumps) and a threefold improvement in spatial resolution. This implies that a magnification factor of ${\rm \mu}=10$ is sufficient to consistently probe physical scales of the order of ${\rm R}_{\rm e}=6\,$pc with MICADO. 

We note that the sample of lensed clumps shown in Fig.~\ref{fig:resk} may not be fully representative of the clump population at cosmic noon. Since strong lensing is a relatively rare phenomenon, lensed galaxies are likely biased toward lower-mass systems and may therefore fail to include the most massive clumps. A significant population of very massive, non-lensed clumps at cosmic noon has recently been identified in JWST/NIRCam observations \citep[see especially Figure 17 of][]{Kalita2025a} with stellar masses reaching $10^{8}-10^{9}\,M_{\odot}$ and sizes up to a few kpc. These massive clumps represent ideal targets for   MICADO, which will have the capability to resolve their internal substructures and detect sub-clumps brighter than the limits shown in Figure~\ref{fig:resk}, down to spatial scales of a few tens of pc. Naturally, compact clumps are expected to be easier to detect than more extended structures with lower surface brightness. The extent to which large clumps can be resolved will depend on the number and size distribution of their substructures, properties that can ultimately only be constrained observationally.
 
We emphasise that the SCAO results are affected by the simulated observing conditions, which have been kept as conservative as possible (i.e. moderate AO reference star brightness, at 30" distance from the target). 
Comparing the SCAO and MCAO modes, we find that the latter yields a gain of approximately $2$ mag relative to the former. 
For compact clumps, ${\rm M_{UV}}=-17\,{\rm mag}$ is reached in the SCAO case; this limit is extended to ${\rm M_{UV}}=-15\,{\rm mag}$ in the MCAO case.
This result is due to the difference in the SR between the two considered PSFs. 
The net effect is that, while for SR$=12\%$ the bright $4\%$ of the considered sample is detectable, the observable fraction rises to $20\%$ for SR$=61\%$. Therefore, for K band observations of non-lensed systems at z$\sim2$, MICADO is unlikely to reach sufficient depth to probe the faint end of the considered clump sample. It is worth noting, however, that a 3 hour integration, leveraging MICADO spatial resolution, is already sufficient to overcome current observational limitations and establish a uniform baseline connecting lensed and unlensed samples.

\subsection{Narrow-band Observations}\label{sec:narrowband}
\begin{figure}
    \centering
    \includegraphics[width=0.9\linewidth]{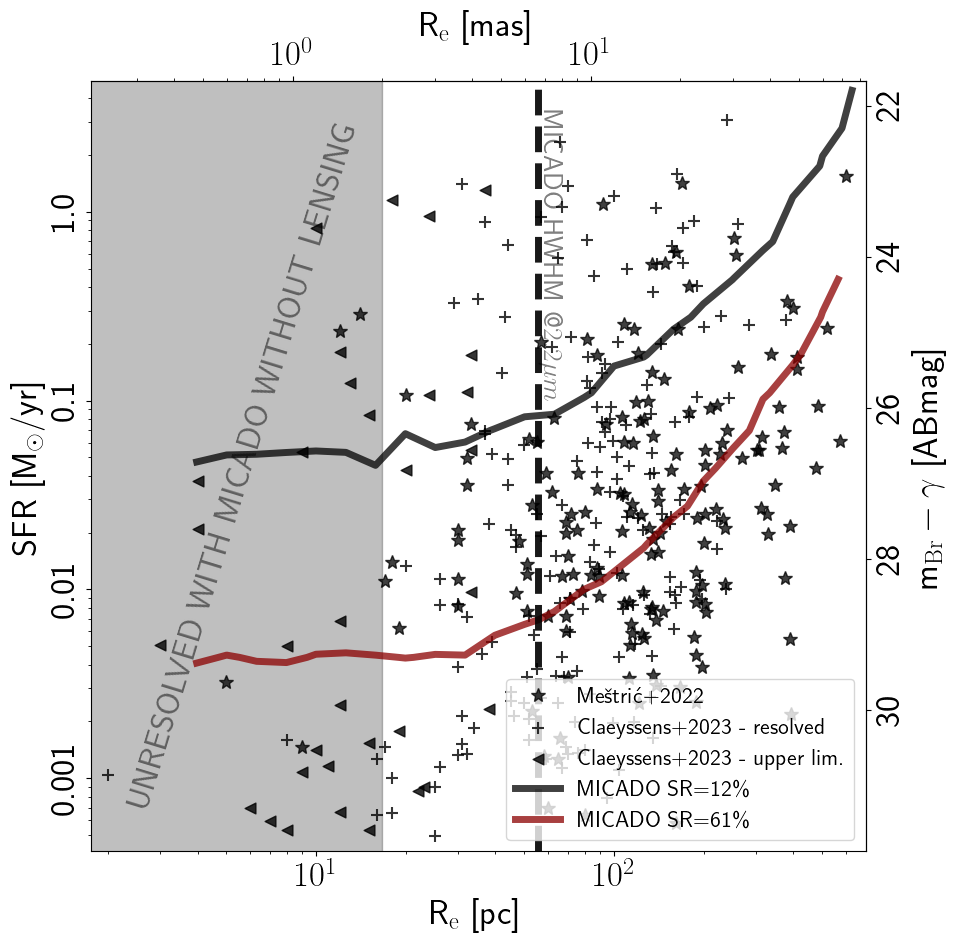}
    \caption{Observability for z$=2.31$, non-lensed clumps using ${\rm Br-\gamma}$ narrow-band filter. Both SCAO and MCAO cases have been considered, $3$ hr total exposure time has been simulated. As in Figure~\ref{fig:resk}, the grey-shaded area highlights the threshold below which clumps cannot be resolved with MICADO, and the dashed vertical line marks the size of MICADO PSF. The SFRs for the same clumps of Figure~\ref{fig:resk} are used as a reference.}
    \label{fig:resha}
\end{figure}
The MICADO ${\rm Br}-\gamma$ filter has a width of about $289$\AA. As a consequence, it can sample the ${\rm H}{\alpha}$ emission in the redshift range ${\rm z}\in[2.29,2.33]$. 
Figure \ref{fig:resha} shows the level of SFR (derived as detailed in Sec.~\ref{sec:host_gal}) detectable as a function of the size of the clumps at an SNR of $5$, keeping the total exposure time of $3\,$hr. Using the MICADO SCAO $30\arcsec$ off-axis PSF (black curve), the detectable SFRs range from $0.5$ to $10\,{\rm M}_{\odot}/{\rm yr}$, for clump sizes between 17 and 600 pc. Also in this case, adopting the MCAO PSF would yield a sensible gain in performance. In terms of SNR, the sensitivity is enhanced by an order of magnitude.

For completeness, we also report in Figure~\ref{fig:resha} the SFRs from the same sample of Figure~\ref{fig:resk}.
Due to their intrinsic luminosity in ${\rm H}{\alpha}$, it can be noticed that the total number of observable targets, in both SCAO and MCAO cases, is higher than what results from broad-band observations. Specifically, almost $20\%$ and more than $50\%$ of the considered sample is detectable in the SCAO and MCAO cases, respectively.
These results make narrow-band observations a promising avenue, even for early MICADO exploitation.  

\section{Discussion}\label{sec:discussion}
\subsection{Clump detectability} \label{sec:discussion:clump_det}
The results in Section~\ref{sec:res} highlight MICADO's effectiveness in characterising non-lensed clumps located at $z\sim2$. 

Referring to the de-lensed galaxy samples of \citet{Mestric2022} and \citet{Claeyssens2023}, we find that the brightest clumps would be observable in broad-band with SR$=12\%$. We stress that this value of SR can be reached in the K band in a region that extends $\sim 0.8{\rm arcmin}^2$ around the AO reference, hence in an area that roughly corresponds to the entire FoV of MICADO. If we require a PSF with SR$\geq61\%$, the number of observable clumps could be increased by a factor $\sim$5 (see Figures~\ref{fig:resk}~and~\ref{fig:resha}). It is important to note, however, that the spatial resolution will be comparable in both SCAO and MCAO cases and will outperform JWST, offering a significant improvement in size characterisation. In non-lensed systems, the spatial resolution of JWST allows, in K band, the characterisation of clump sizes larger than ${\rm R}_{\rm e}=100\,$pc (e.g. \citealt{Kalita2025a}), $20\%$ clusters less than MICADO at the same depth as the MCAO case.

Narrow-band observations represent an attractive development. These observations will benefit from the enhanced intrinsic brightness of the ${\rm H}{\rm \alpha}$ emission line from the clumps. If a target with a redshift in a narrow range ($2.29\lesssim{\rm z}\lesssim2.33$) is identified, more than $10\%$ of the reference sample can be observed even with SR$=12\%$, which is more than double what can be achieved, in the same conditions, with broad-band observations.

\subsection{Clump substructures}
\begin{figure*}
    \centering
    \includegraphics[trim={1cm 0 2cm 1cm},clip,width=0.9\columnwidth]{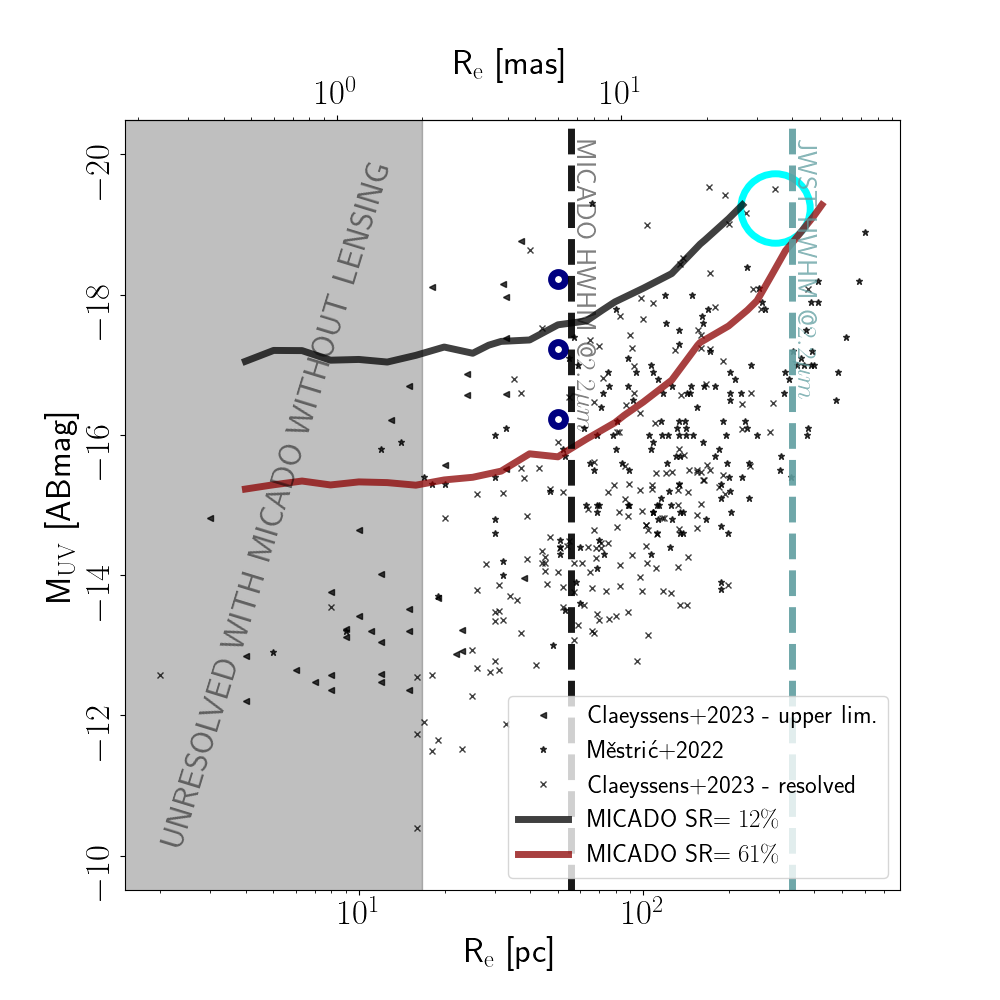}
    \includegraphics[width=0.9\columnwidth]{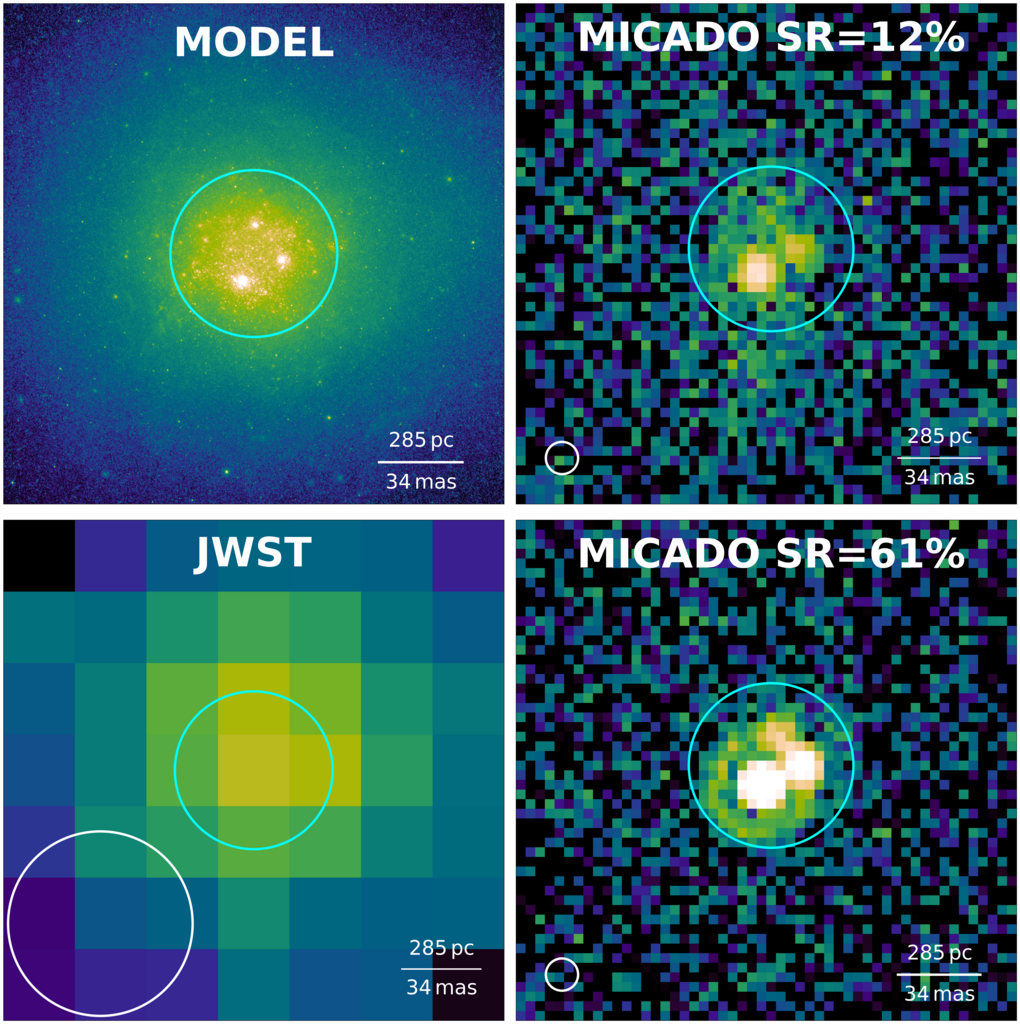}
    \caption{Simulated clump sub-structure observability at z=2.31. {\em Right panel:} The model image of a single, extended, massive clump is shown in the upper-left sub-panel. Its size is marked by the cyan circle, having a radius equal to R$_{\rm eff}=285\,$pc ($=34\,{\rm mas@z}=2.31$). In this case, we assume most of its light is distributed among $3$ sub-clumps of different luminosities. Additional compact, faint structures are included along with an extended component. The right sub-panels show the simulated MICADO observations considering the SR$=12\%$ and SR$=61\%$ cases (upper and lower sub-panel, respectively). The $3$ main components are spatially resolved. A simulated JWST observation of the same target is shown in the bottom-left sub-panel for comparison. In the latter, the clump is not spatially resolved. For reference, white circles in the bottom-left corners of the middle and right panels have a diameter equal to the PSF FWHM of MICADO and JWST, respectively. {\em Left panel:} As in Figure~\ref{fig:resk}, highlighting size and luminosity for the simulated star-forming region (cyan circle), along with those of its $3$ main sub-components (blue circles).}
    \label{fig:sim_csub}
\end{figure*}
To further highlight the substantial scientific value that MICADO observations will deliver, we discuss the characterisation of potential sub-clumps within compact star-forming regions. 
Specifically, we modelled a hypothetical compact star-forming region at z$=2.31$ consisting of an extended component with three point-like sub-clumps located near its centre (Figure~\ref{fig:sim_csub}, right panel, upper-left subpanel). This region has an effective radius consistent with the average sizes of the \citet{Kalita2025b} sample: R$_{\rm eff} = 285$pc (corresponding to $34\,{\rm mas@z}=2.31$, indicated by the azure circles in all panels). Its three bright sub-clumps are separated by $150-190\,$pc ($18-23\,$mas at z$=2.31$). To construct this model, we leveraged HST observations of NGC$\,1705$ (${\rm D}_{\rm NGC\,1705}=5.1\,$Mpc) released as part of the LEGUS project \citep{Calzetti2015_legus} as they provide the required spatial resolution to model fine structural details at z$=2.31$. To simulate the sub-clumps, we combined three copies of the same image, each slightly off-centre and rotated by a different angle.
We assigned a luminosity to each image component such that the sub-clumps have magnitudes of m$_{\rm K}=27\,{\rm mag}$, m$_{\rm K}=28\,{\rm mag}$ and m$_{\rm K}=29\,{\rm mag}$ respectively. These values are consistent with the detection limits discussed in the previous sections, resulting in an integrated magnitude of m$_{\rm K}=26\,{\rm mag}$ for the entire star-forming region. The finalised model image is shown in the right panel of Figure~\ref{fig:sim_csub} (top-left inset). For clarity, the positions of the simulated star-forming region and its individual sub-components are plotted in the same size-luminosity plane of Figure~\ref{fig:resk} (left panel of Figure~\ref{fig:sim_csub}, azure and blue circles). 

The resulting simulated MICADO MCAO observation (Figure~\ref{fig:sim_csub}, right panel, bottom-right inset) clearly shows that all three sub-clumps are detectable and spatially resolved. The PSF size, indicated by the white circle in the bottom-left corner, is approximately $1.5$ times smaller than the apparent separation between substructures.
In contrast, in the SCAO configuration in Figure~\ref{fig:sim_csub}, right panel, top-right inset), only one of the three sub-clumps achieves a SNR>5.
For comparison, in Figure~\ref{fig:sim_csub} we also show simulated JWST observations (right panel, bottom-left inset) for the same total exposure time. In this case, the compact region is completely blended and remains spatially unresolved. Consequently, while accurately modelling the JWST PSF may yield an estimate of the global effective radius, the spatial information regarding its internal sub-clumps is entirely lost.

\subsection{Clump target selection}
One issue that MICADO will face is the selection of the targets.  
Based on the number density of star-forming galaxies with $1.4\leq{\rm z}\leq2.5$ \citep[e.g.][]{2014A&A...571A..99S}, we expect, on average, 
$16$ $z\sim2$ star-forming galaxies with integrated magnitude ${\rm m}_{\rm K}<24$~mag to fall within a MICADO pointing. This number grows to $90$ if the sample is extended to galaxies with ${\rm m}_{\rm K}<27$~mag. 
These numbers are consistent with what has been observed in the GOODS-South field (as part of CANDELS, see \citealt{Grogin2011,Koekemoer2011}). By querying the Gaia DR3 Synthetic Photometric Catalogue (GSPC), in search for suitable SCAO AO reference sources, we counted $8$ stars with m$_{\rm R}<16\,$mag that fall in the considered field. For each of them, we measured the number of galaxies with $1.4<{\rm z}<2.5$ and ${\rm m}_{\rm K}<24\,$mag that are less than $30\,$arcsec away \citep{Guo2013}. It results that, on average, $15$ galaxies per star are found within a MICADO pointing, confirming previous estimations.
Since only $60\%$ of $z\sim2$ star-forming galaxies are clumpy \citep{Shibuya2016}, the number of suitable targets per pointing  drops to $10$ ($54$) for ${\rm K}<24$~mag (${\rm K}<27$~mag).

Taking as a reference the average number of detected clumps per galaxy from \citet{Claeyssens2023}, about $15$ clumps per galaxy for $1.4<{\rm z}<2.5$, it is possible to estimate the expected number of clumps per MICADO pointing. From the results presented in Sec. \ref{sec:discussion:clump_det}, in the MICADO SCAO case, statistically the brightest $4\%$ of the clump population is detectable. This results in $6$ detectable clumps per MICADO pointing in broad-band if SR$=12\%$ is considered ($15$ clumps in narrow-band). If the SR$=61\%$ case is considered, the number of observed clumps per MICADO pointing in broad-band increases to $30$. These estimates assume that clump populations derived from strongly lensed systems are representative of the general population at similar redshifts. Future high-resolution observations of blank-field galaxies will provide an important opportunity to independently assess and refine this picture.

Finally, the bright AO reference star is likely saturated during exposures optimised for this specific science case. A possible observing strategy to avoid saturation would be to place the AO reference star in the gaps between detectors, or slightly outside of the scientific image. In the latter case, this would drastically reduce the effective area of the MICADO FoV, where the achieved SR enables proper characterisation of star-forming clumps (e.g. luminosity and size measurements), decreasing the number of suitable targets. Another option, more preferable in terms of constructing efficient dithering patterns, is to read the detector faster just on the SCAO AO reference star. For the MICADO imaging mode, in fact, there will be the option (already tested in laboratory conditions) of flagging bright stars across the detector mosaic and, in correspondence with such bright objects, reading the detector multiple times\footnote{These regions are reset multiple times during the readout of the whole frame.} to prevent saturation.

\subsection{Impact of the reconstructed PSF on size measurement and magnitude determination}\label{sec:scieval}
The analysis above has been based on the knowledge of the PSF that characterises the images. Nonetheless, obtaining   
a good PSF template is not trivial, especially in the case of extragalactic fields, at high Galactic latitudes, where the number of bright stars is very low, as discussed in the previous section. 

One of the deliverables of the MICADO consortium for the end user will be a PSF-R service that provides template PSFs for each observation without relying on the scientific frames \citep{2020SPIE11448E..37S,2022SPIE12185E..41G}. Instead, the reconstruction uses AO telemetry, complemented by atmospheric data and careful calibration of instrumental effects. For SCAO observations, the proposed PSF-R method in the on-axis case \citep{2018JATIS...4d9003W} has been validated on real data coming from current AO facilities \citep{2022JATIS...8c8003S} and extended to the off-axis case, adopting a tomographic method \citep{2023JOSAA..40.1382W}. We refer the reader to those works for more details. Briefly, the method estimates the phase difference for an off-axis direction by combining incoming wavefront measurements towards the AO reference source with external wind vectors and atmospheric turbulence measurements. The adopted MICADO SCAO PSF results from a simulation under fixed atmospheric conditions \citep[generated using COMPASS,][]{2014SPIE.9148E..6OG}; it is hence obtained in a relatively controlled environment.

\begin{figure}
    \centering
    \includegraphics[width=0.9\linewidth]{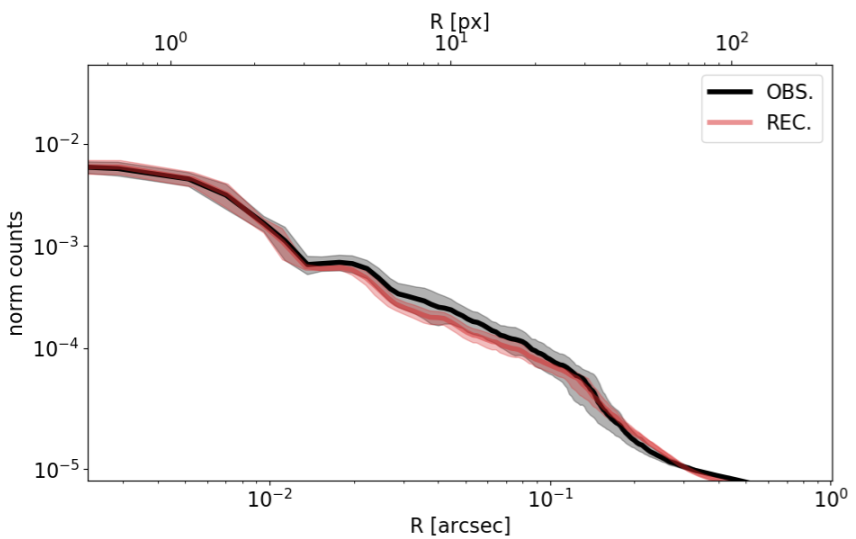}
    \caption{Comparison between the adopted MICADO SCAO $30"$ off-axis PSF (black solid line) and the reconstructed one (in red). The radial profiles are shown, and a log scale is used for all axes. The shaded regions show the measured standard deviation in concentric annuli.}
    \label{fig:psfr}
\end{figure}
The result of the PSF-R in the case of the MICADO SCAO $30\arcsec$ off-axis, K band PSF, is shown in Figure~\ref{fig:psfr}. The radial profiles show striking agreement between the original and the reconstructed PSF, as quantified in Table~\ref{tab:psfr}. A classic PSF metric is used to assess the performance of the reconstruction, taking into account SR, full-width at half maximum (FWHM), encircled energy of the core (${\rm EE}_{\rm C}$), and half-light radius (${\rm R}_{50}$).

\begin{table}
    \centering
    \caption{\label{tab:psfr} Properties of the reconstructed MICADO SCAO PSF. It refers to an off-axis distance of $30"$, in the K band. The relative difference between the reconstruction and the original MICADO SCAO PSF is reported in the last column. The listed parameters include SR, full-width at half maximum (FWHM), encircled energy of the core (${\rm EE}_{\rm C}$), and half-light radius (${\rm R}_{50}$).}
    \begin{tabular}{l|r|r}
    Parameter          & Value        & Rec. Rel. Err. $[\%]$\\
    \hline
    SR                 & $11.6\%$     & -3  \\
    FWHM               & $12.6$ mas   & -6  \\
    ${\rm EE}_{\rm C}$ & $7\%$        & +5  \\
    ${\rm R}_{50}$     & $129$ mas    & +25 \\  
    \end{tabular}
\end{table}
The achieved precision in SR, FWHM, and EE$_{\rm C}$ is around $5\%$. 

Following \cite{2022JATIS...8c8003S}, we assess the impact of using the reconstructed PSF for the characterisation of the clumps, including their magnitude and size. Specifically, we analyse the 
$144\times120$ clumps in the simulated broad-band observations by running GALFIT with perturbed initial guess values for the parameters. We assume that all the clumps are circular, and we apply a flat distribution to perturb the integrated magnitudes within $[-0.5, +0.5]$ mag and the effective radii within a $[-10\%, +10\%]$ range. We use the exact values for all the other quantities (the S\'ersic index, the eccentricity, and the position angle), 
as well as for the properties of the host galaxy, while keeping the clump position and sky value variable.
We then perform the measurements twice, once using the same MICADO SCAO $30\arcsec$ off-axis PSF used for the simulated observations (the control measurement), and another time using the resulting reconstructed PSF (the actual measurement). This enables us to disentangle the effect of PSF reconstruction alone from the intrinsic bias of the measuring process.

\begin{figure*}
    \centering
    \includegraphics[width=0.99\columnwidth]{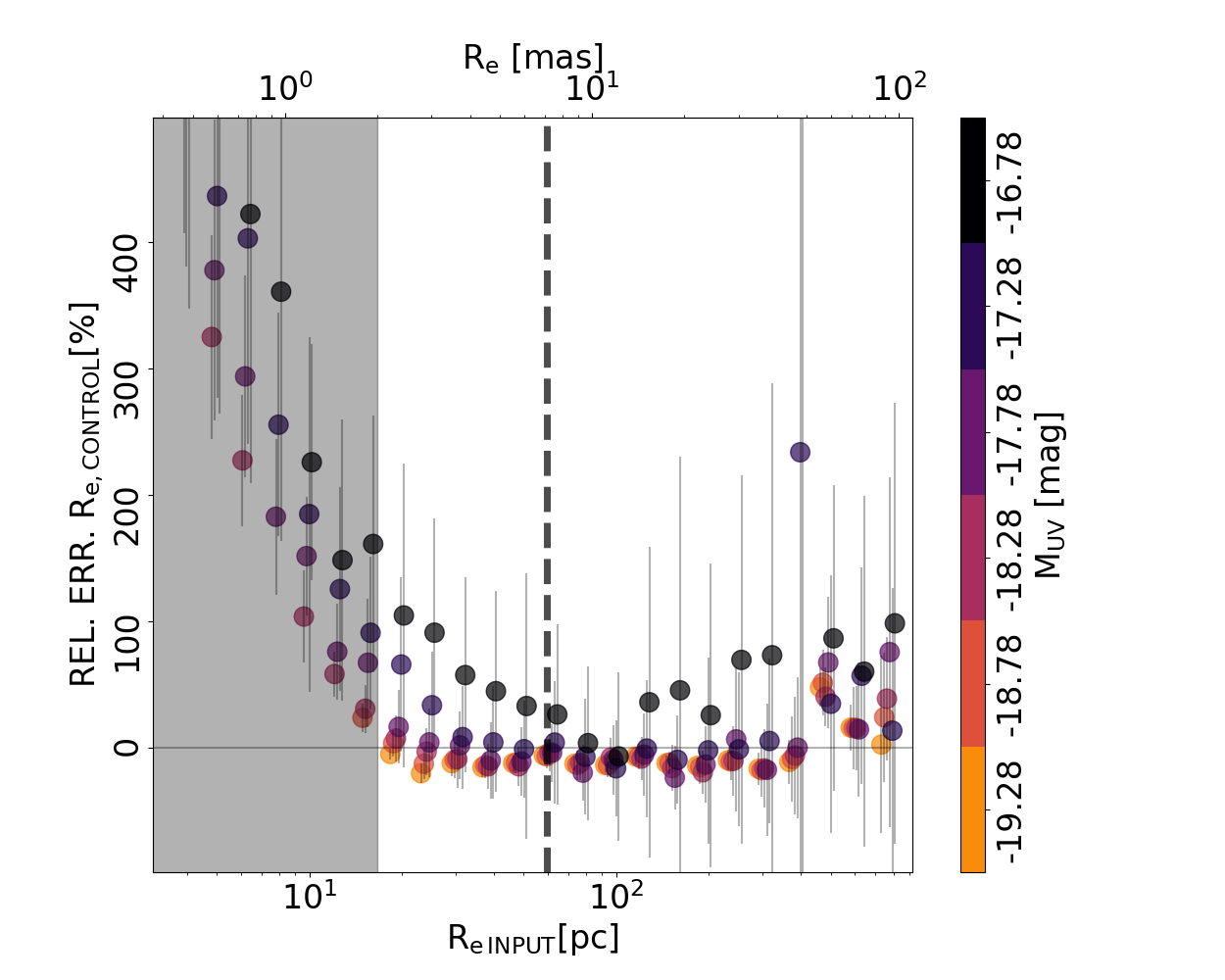}
    \includegraphics[width=0.99\columnwidth]{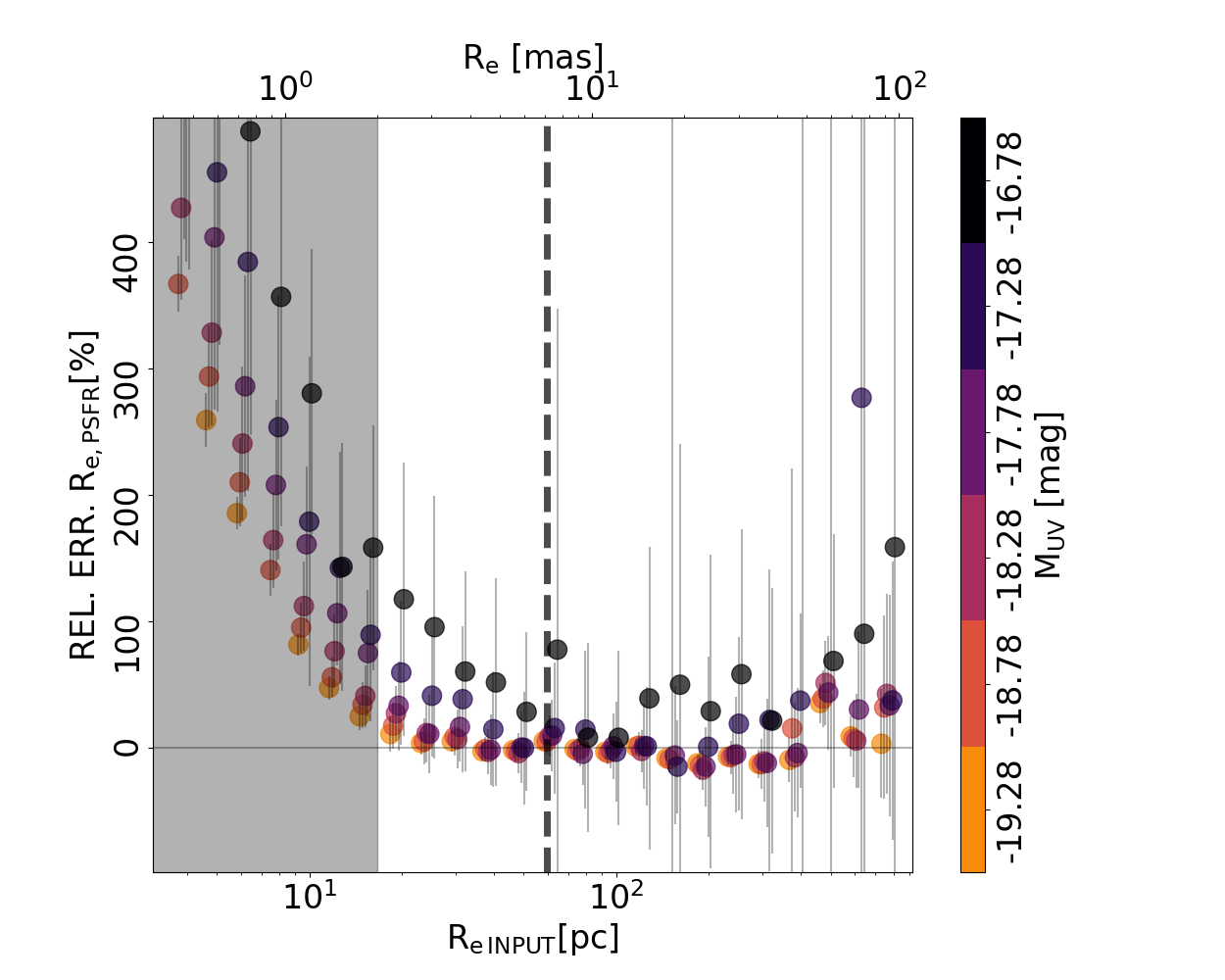} \\
    \includegraphics[width=0.99\columnwidth]{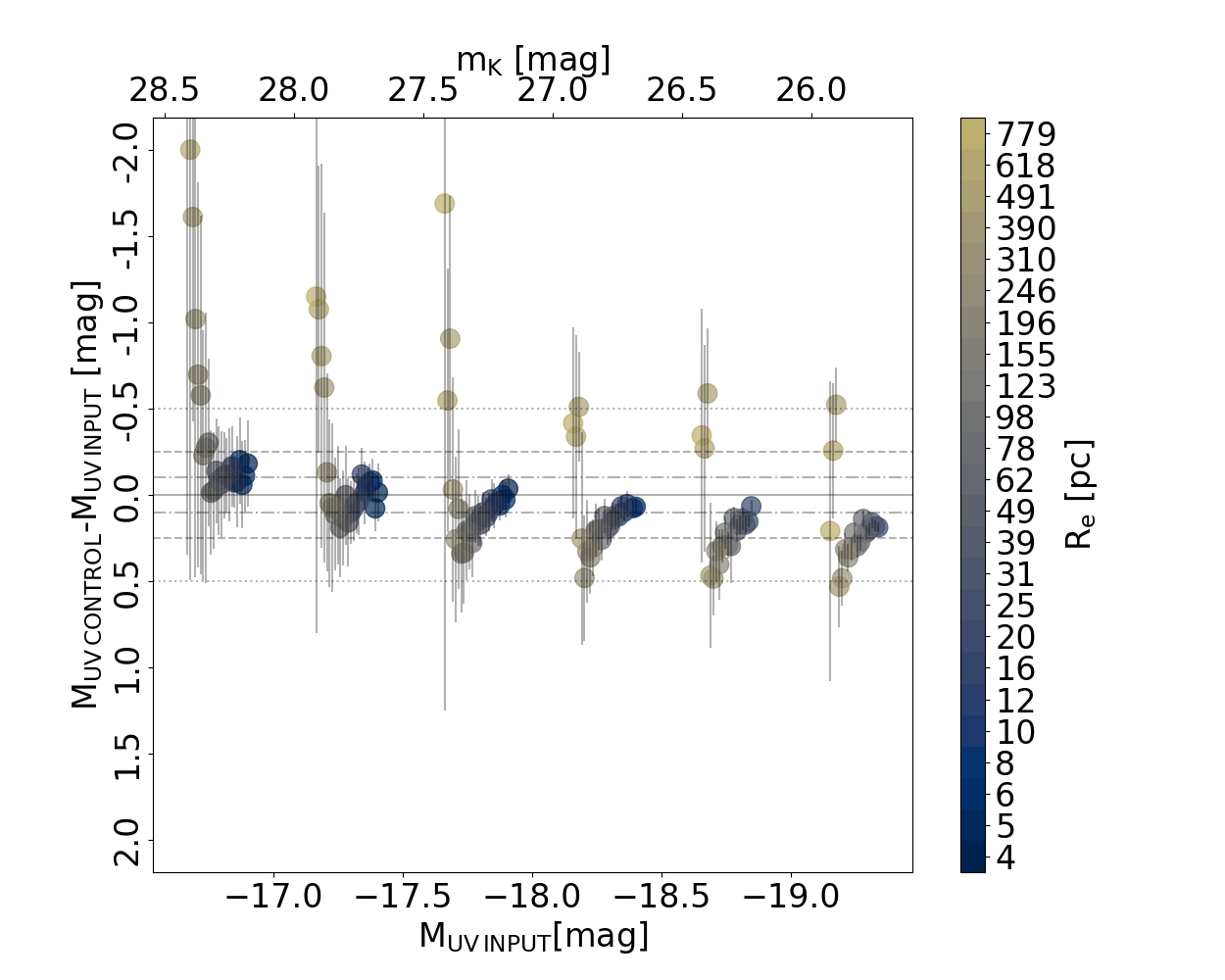}
    \includegraphics[width=0.99\columnwidth]{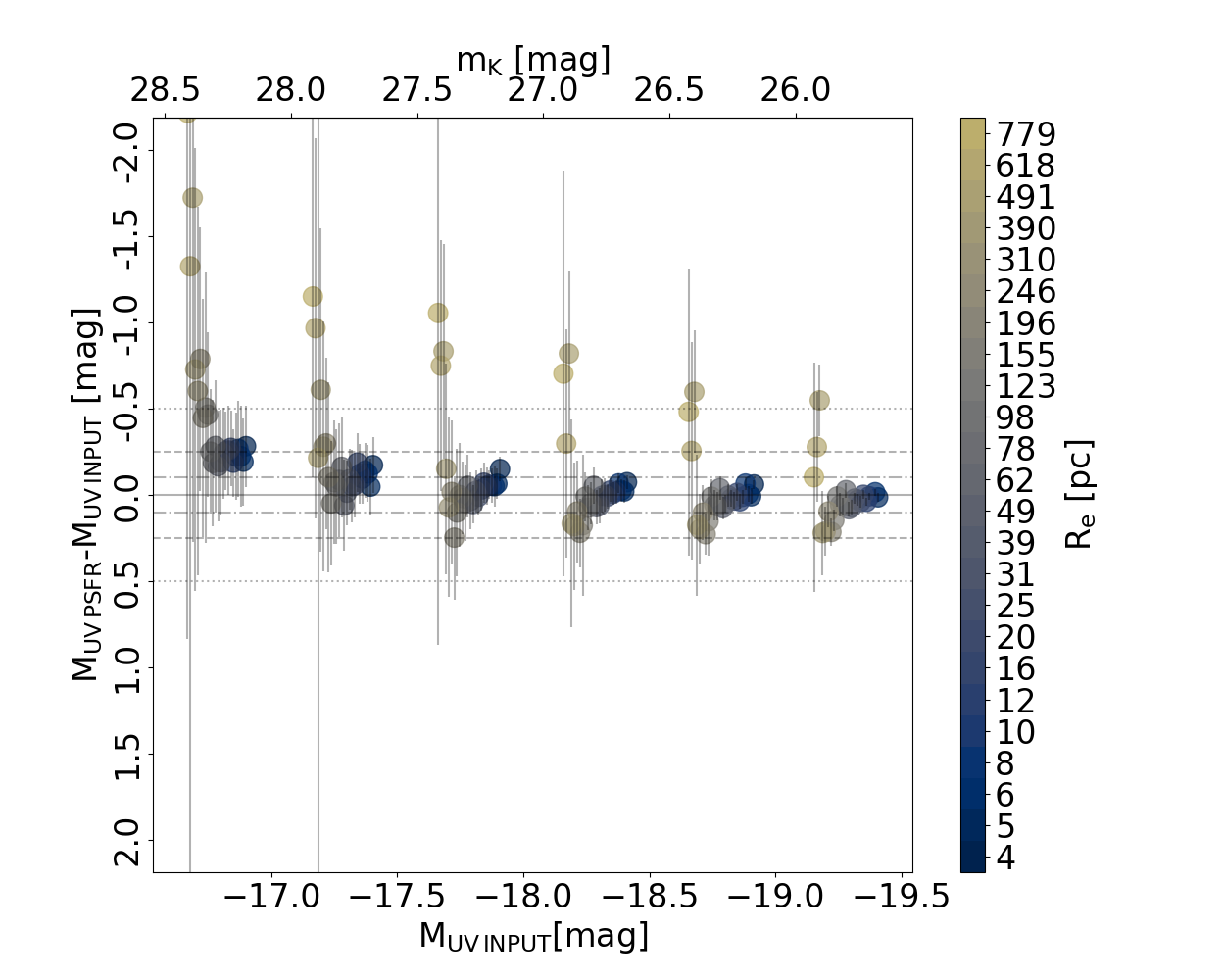} \\
    \caption{Capability of recovering clumps’ sizes and magnitudes. {\em Left panels:} Results for the control measurement, where the same SCAO PSF used in the simulated observations was provided to GALFIT to measure the properties
    of the simulated clumps. {\em Right panels:} As in the left panels, but in this case, the reconstructed PSF (rPSF) is used as input for GALFIT. 
    {\em Top panels:} The relative error on the effective radius of the clusters is shown as a function of the input values. Scales are logarithmic, points are colour-coded based on the clump integrated magnitude, and for each ${\rm R}_{\rm e}$ value, an arbitrary shift has been introduced for clarity. For completeness, the shaded region and the dashed vertical line are the same as in Figure~\ref{fig:resk}. {\em Bottom panels:} The absolute error on the integrated magnitude determination for each cluster. The points are colour-coded for ${\rm R}_{\rm e}$ and again, an arbitrary shift has been introduced for clarity at each ${\rm M}_{\rm UV}$ value. Differences of $0.1$, $0.25$, $0.5$ mag are marked with horizontal dot-dashed, dashed, and dotted lines, respectively.}
    \label{fig:psfr_meas}
\end{figure*}
 For each simulated clump, in each set, we compute the average values of the recovered size and luminosity.
 In particular, we defined as recovered the simulated clumps with at least $12$ valid measurements, i.e. $10\%$ of the total simulations for each clump size-luminosity combination. 
 We also estimate the internal dispersion of each average by computing the standard deviation among the valid measurements. 

Figure~\ref{fig:psfr_meas} shows the results of our analysis: the top panels show the comparison between the input and measured ${\rm R}{\rm e}$ values for the control PSF (left) and the reconstructed one (right), while the bottom panels show the same for the integrated magnitudes.  

In general, a good match between the input values and control measurement is found, with no significant trends based on the clump's luminosity. However, for clump sizes of ${\rm R}_{\rm e}\lesssim2$ mas ($0.5$ MICADO pixels in wide mode), the method loses precision. This region, where clumps are unresolved, is highlighted in grey in the figure. The same threshold is used in Figures~\ref{fig:resk},~\ref{fig:resha}~and~\ref{fig:sim_csub}. 
For reference, the MICADO HWHM is shown as a vertical dashed line. A detailed knowledge of the PSF is crucial for accurately measuring clump sizes in this region between the shaded area and the dashed line. For extended clumps, sampling light at MICADO’s fine scale ($4\,$mas/pixel) causes the wings of the luminosity profile (assumed Gaussian) to drop rapidly below the background noise floor. Consequently, only the central core achieves S/N$>5$, leading GALFIT to systematically overestimate both the size and total flux (recovering brighter magnitudes). We note that such large structures (${\rm R}_{\rm e} \gtrsim 300\,$pc) may be intrinsically rare; their higher prevalence in unlensed, blank field studies likely stems from spatial blending of unresolved sub-clumps, a hypothesis MICADO is ideally suited to test.

\begin{figure*}
   \centering
   \includegraphics[width=0.99\columnwidth]{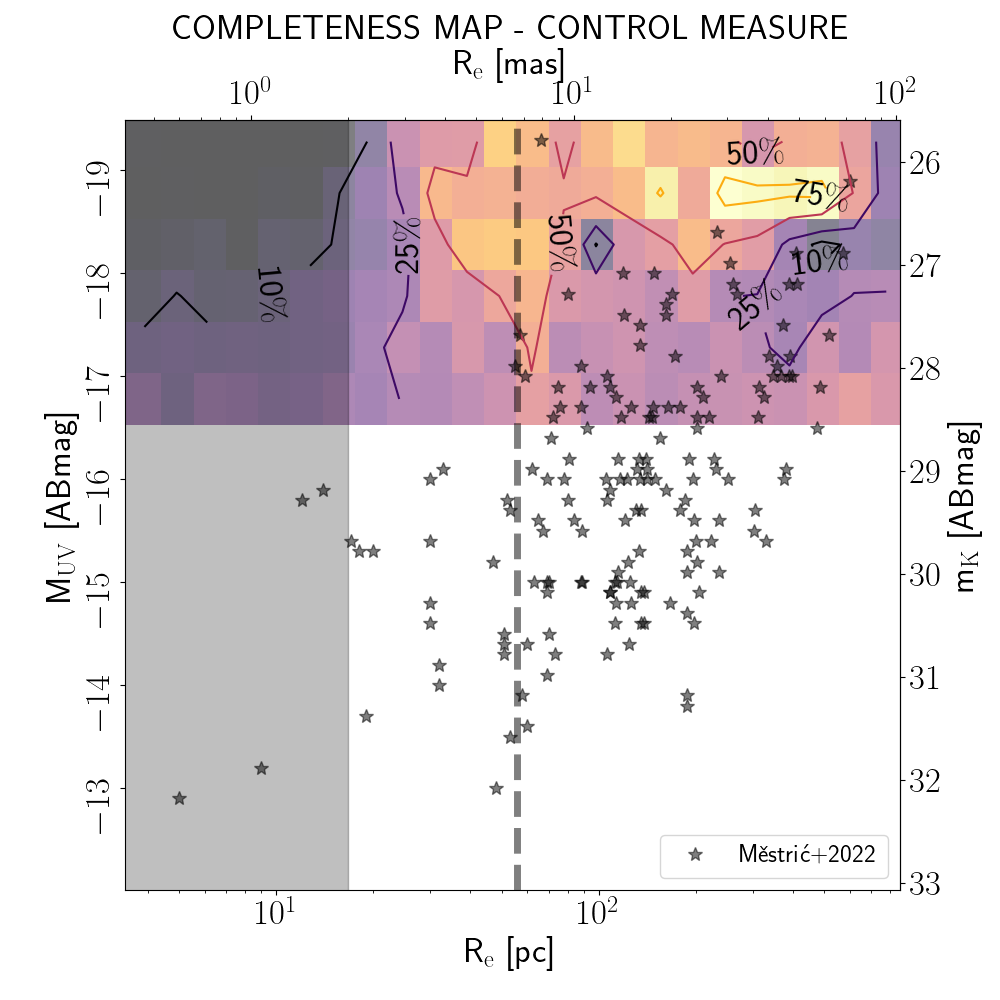}
   \includegraphics[width=0.99\columnwidth]{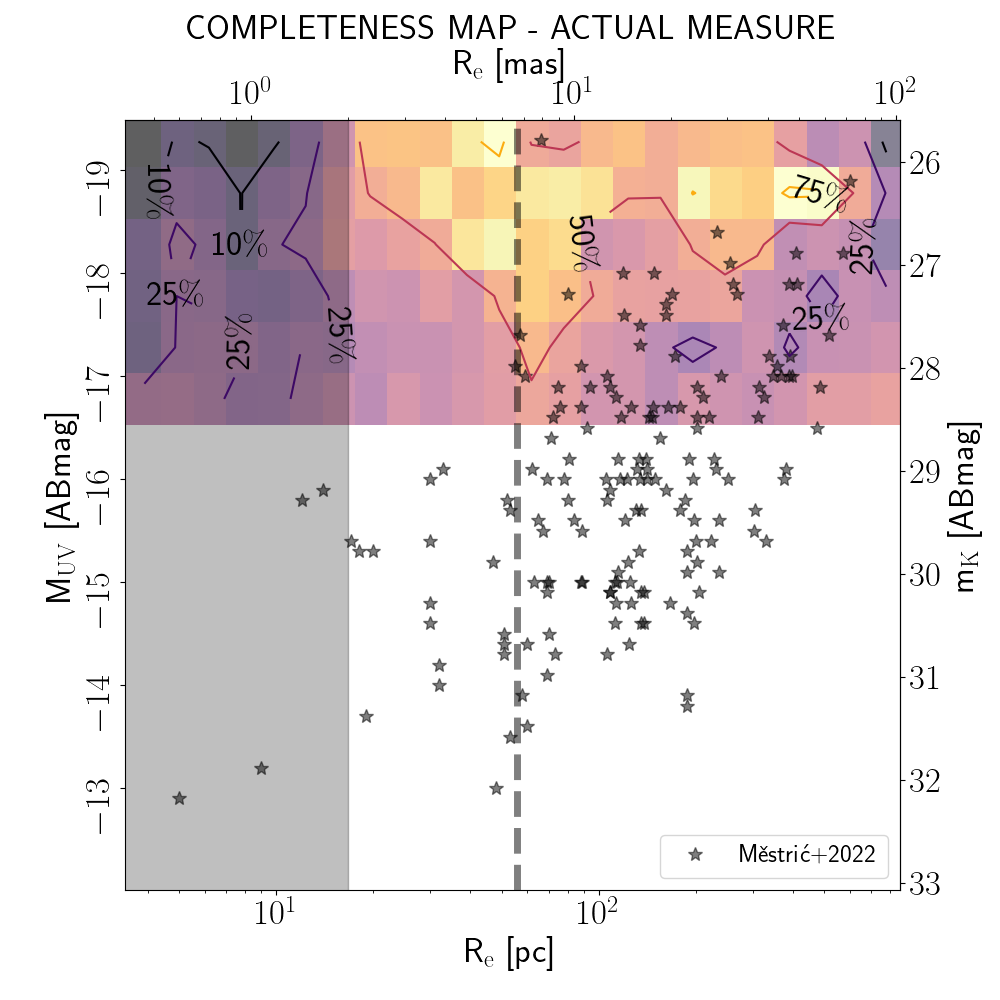}
   \caption{Completeness associated with the measuring process for the considered SCAO case. Results for the control measurement (left) and the actual measurement (right) are shown. The colour code of the map and the associated completeness levels are the same and represent the fraction of valid measurements, with respect to the total number of simulated clumps, for each clump luminosity and size. The sample of \citet{Mestric2022} is overplotted as a reference along with the MICADO HWHM (black dotted vertical line) and the same shaded region as in Figure \ref{fig:resk}. The colour scale used is proportional to the completeness level, indicated by the contour lines that mark $10\%$, $25\%$, $50\%$ and $75\%$ completeness.}
    \label{fig:completeness}
\end{figure*}
The completeness values associated with the measuring process are shown in Figure \ref{fig:completeness} for both the control measurement (left panel) and the actual measurement (right panel).
For what concerns the control measurement, $23$ individual combinations of magnitude and size of the simulated clumps ($16\%$ of total cases), fewer than $12$ valid measures have been registered. These are all associated with bright and compact (i.e. unresolved) objects. These objects behave like point sources, making their size measurement unrealistic. 
For half the simulated cases, at least $40$ valid measurements have been obtained ($30\%$ of the $120$ realisations of each simulated configuration). And at least $100$ valid measurements have been obtained in $5\%$ of the simulated configurations.

The right panel of Figure \ref{fig:completeness} shows the results for the actual measurement. 
The map is qualitatively similar to the control measurement, but in this case, $3\%$ of the simulated configurations of clump magnitude and size are associated with less than $12$ valid measurements. In half the cases, at least $50$ valid measurements are registered, and more than $100$ valid measurements are obtained for about $10\%$ of the simulated configurations. Rejected measurements are again mainly associated with bright, extremely compact sources. 

It is important to note that the input values for GALFIT were randomly perturbed around the correct values for both the control and actual measurements, which allows us to isolate the effects of using a different PSF between the two tests.

As expected, using the reconstructed PSF has little effect on clumps larger than the PSF FWHM. 
However, even for clumps in the range $2\,{\rm mas}<{\rm R}_{\rm e}<6.7\,{\rm mas}\,({\rm HWHM})$, the reconstructed PSF enables successful estimation of the clump sizes.

A similar trend is observed for the integrated magnitudes of the clumps, where the primary effect of using the reconstructed PSF is seen for the fainter, larger clumps.

\section{Conclusion}\label{sec:wup}
While JWST has provided unprecedented insights into compact star-forming clumps in high-redshift galaxies, its ability to characterise objects smaller than 100 pc is limited. Detailed investigations of these sub-100 pc objects typically rely on strong gravitational lensing. This constraint restricts the analysis to the surroundings of massive galaxy clusters, while introducing significant systematic uncertainties stemming from the modelling of the cluster mass distribution.
As a result, the vast majority of compact clumps remain inaccessible at the spatial scales required to understand their internal properties and formation mechanisms. In this context, the upcoming operations of the ELT and its first-light instrument, MICADO, present an excellent opportunity to overcome these limitations and probe non-lensed clumps at parsec scales.  

In this paper, we have investigated the performance of MICADO in characterising star-forming clumps in $z\sim2$, non-lensed galaxies.
Through a series of idealised (but realistic) simulations, we explored the detectability and accuracy in determining the magnitude and physical size of these clumps. We simulated observations using both the K-band and the narrow Br-$\gamma$ filter. In the K-band, our goal was to assess the quality of the MICADO PSF at this wavelength and determine whether it meets the required specifications. The advantages of this wavelength regime include the ability to achieve a relatively high SR across the entire MICADO FoV and its reduced sensitivity to sky background. For the narrow band filter, our goal was to simulate observations specifically designed to study ${\rm H}{\rm \alpha}$ emitting clumps at z$=2.31$. 
 
The SCAO approach provides near-diffraction-limited observations in the near-IR in the immediate vicinity of the bright point source used as the AO reference, while its performance rapidly degrades as one moves away from the reference star, with this effect being more pronounced at shorter wavelengths. Under typical atmospheric conditions, in the K band, MICADO will deliver high SR PSFs of approximately $\sim60\%$ in the immediate vicinity of the bright  (${\rm m}_{\rm R}\leq14$mag) point source. At $30\arcsec$  from the AO reference, the expected PSF SR drops to about $10\%$. 

We demonstrated that, provided a suitable AO reference star is available,  the entire MICADO FoV ($50.5\arcsec\times50.5\arcsec$) can be effectively used to pursue the selected science case. Taking into account the number density of star-forming galaxies at $1.4<{\rm z }<2.5$, their brightness and clumpiness  \citep{2014A&A...571A..99S,Shibuya2016}, we estimated that a MICADO pointing could successfully observe about 10 clumpy galaxies at a time. 

Our analysis demonstrates that MICADO will routinely detect and characterise clumps as small as 20 pc in non-lensed galaxies. These performances will further improve when MORFEO complements the SCAO mode, providing a more uniform AO correction across the FoV. 
With 3 hours of observations and assuming a MICADO SCAO PSF with SR$=12\%$, we expect to observe clumps up to sizes of ${
\rm R_{e}}\sim20\,{\rm pc}$ and absolute UV rest-frame magnitude ${\rm M_{UV}}\sim -17{\rm mag}$. 
This is roughly $4\%$ of the combined sample of \citet{Mestric2022} and \citet{Claeyssens2023}, which are representative of the state-of-the-art results from JWST observations of gravitationally lensed systems.
This percentage of detected clumps increases to 10\% when the narrow band filter is used and hence when we focus on ${\rm H}{\rm \alpha}$ emission, with a SR$=12\%$. In case SR=61\% will be reached, the number of observable clumps would be increased by a factor of 5 in both broad- and narrow-band observations.

To conclude, in this paper, we have demonstrated that the exceptional spatial resolution of MICADO, coupled with its unprecedented sensitivity for a ground-based facility, will enable the characterisation of clumps on scales of tens of pc in non-lensed systems. The presence of strong gravitational lensing would push these limitations to even smaller scales and fainter magnitudes, proportionally.
\section*{Acknowledgements}
We thank the anonymous referee for the constructive and careful review that helped improve clarity and overall quality of the paper.
MS and AG acknowledge partial support from Bando Ricerca Fondamentale INAF 2023 (Ob. Fu. 1.05.23.04.05, P.I. Simioni). AZ acknowledges support from the European Union – NextGeneration EU within PRIN 2022 project n.20229YBSAN - Globular clusters in cosmological simulations and in lensed fields: from their birth to the present epoch and the INAF Minigrant ‘Clumps at cosmological distance: revealing their formation, nature, and evolution (Ob. Fu. 1.05.23.04.01). LS acknowledges the financial support from the PhD grant funded on PNRR Funds Notice No. 3264 28-12-2021 PNRR M4C2 Reference IR0000034 STILES Investment 3.1 CUP C33C22000640006. EVal acknowledges the Excellence Cluster ORIGINS Funded by the Deutsche Forschungsgemeinschaft (DFG, German Research Foundation) under Germany’s Excellence Strategy – EXC-2094-390783311.
This work has made use of data from the European Space Agency (ESA) mission Gaia (https://www.cosmos.esa.int/gaia), processed by the Gaia Data Processing and Analysis Consortium (DPAC, https://www.cosmos.esa.int/web/gaia/dpac/consortium). Funding for the DPAC has been provided by national institutions, in particular the institutions participating in the Gaia Multilateral Agreement.

\section*{Data Availability}

The data underlying this article will be shared on reasonable request to the corresponding author. 


\balance
\bibliographystyle{mnras}
\bibliography{mc_biblio} 




%
%


\bsp	
\label{lastpage}
\end{document}